\documentclass{article}

\usepackage{arxiv}

\usepackage[utf8]{inputenc} 
\usepackage[T1]{fontenc}    
\usepackage{hyperref}       
\usepackage{xurl}            
\usepackage{booktabs}       
\usepackage{amsfonts}       
\usepackage{nicefrac}       
\usepackage{microtype}      
\usepackage{lipsum}		
\usepackage{graphicx}
\usepackage{natbib}
\usepackage{doi}
\usepackage{multirow}
\usepackage{amsmath}

\title{Tracing the Techno-Supremacy Doctrine: A Critical Discourse Analysis of the AI Executive Elite}

\author{ \href{https://orcid.org/0009-0004-6548-9852}{\includegraphics[scale=0.06]{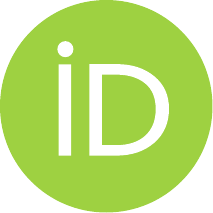}\hspace{1mm}H\'ector P\'erez-Urbina} \\
	University of Cambridge\\
	Cambridge, UK \\
	\texttt{hmp56@cam.ac.uk}
}

\date{}

\renewcommand{\shorttitle}{Tracing the Techno-Supremacy Doctrine}

\hypersetup{
pdftitle={Tracing the Techno-Supremacy Doctrine: A Critical Discourse Analysis of the AI Executive Elite},
pdfsubject={cs.CY},
pdfauthor={H\'ector P\'erez-Urbina},
pdfkeywords={Responsible AI, Critical Discourse Analysis, Science and Technology Studies, Techno-Supremacy Doctrine, AI Executive Elite, Techno-Optimism},
}

\begin{document}
\maketitle

\begin{abstract}
    This paper critically analyzes the discourse of the `AI executive elite,' a group of highly influential individuals shaping the way AI is funded, developed, and deployed worldwide. Its primary objective is to examine the presence and dynamics of the `Techno-Supremacy Doctrine' (TSD), a term coined for this research to describe a belief system characterized by an excessive trust in technology's alleged inherent superiority in solving complex societal problems. This paper presents a novel approach to Critical Discourse Analysis (CDA) that responds to scholarly critiques by integrating quantitative metrics with in-depth qualitative investigations. This methodology is operationalized to analyze 14 texts published by elite members between 2017 and 2025. The findings demonstrate that the elite is not a monolithic bloc but exhibits a broad spectrum of stances. The discourse is highly dynamic, showing a marked polarization and general increase in pro-TSD discourse following the launch of ChatGPT. The analysis identifies key discursive patterns, including a dominant pro-TSD narrative that combines utopian promises with claims of inevitable progress, and the common tactic of acknowledging risks only as a strategic preamble to proposing further technological solutions. This approach to CDA acts as a linguistic `diagnostic toolkit' for identifying TSD's diverse manifestations, from insidious to benign. This paper argues that fostering critical awareness of TSD's discursive patterns is essential for AI practitioners, policymakers, and the general public to actively navigate the future of AI.
\end{abstract}


\section{Introduction}
\label{sec:intro}

The power of technology is undeniable, and its allure seductive, particularly in light of the seemingly ever-growing capabilities of the latest generative Artificial Intelligence models (e.g., ChatGPT). Amid this admittedly awe-inspiring and accelerated development, it is tempting to think of technology as inherently superior at tackling not only mundane tasks but also complex societal issues. This trust in technology's prowess can become excessive and uncritical, constituting the foundation of the \emph{Techno-Supremacy Doctrine} (TSD).

Adhering to TSD can be problematic. It can lead to oversimplifying problems, devaluing alternative approaches, and uncritically accepting technology. For example, TSD can manifest perniciously in education, where the intricate process of learning is often reduced to a quantifiable output amenable to optimization. This mindset posits that systemic issues such as educational disparities can be ``fixed'' by deploying technological tools (e.g., laptops, AI tutors). While the use of AI as an educational aid is promising, there are significant risks, as it may affect cognitive processes essential to genuine learning. Recent research shows that students using Large Language Models (LLMs) for essay writing displayed significantly weaker brain engagement, accumulated a ``cognitive debt,'' and felt a lower sense of ownership over their work compared to control groups \citep{kosmyna_your_2025}. Focusing on the final product (e.g., an essay) over the cognitive effort can narrow educational goals to ``teaching to the test,'' substituting deep understanding and expert human mentorship for the optimization of simplistic metrics \citep[p.~94]{toyama_geek_2015}.

TSD's risks are significantly amplified when it is adopted by people with the power to direct how AI is funded, developed, and deployed on a global scale. Through their discourse, this \emph{AI executive elite} uses public platforms to shape narratives, influence policy, and define what is considered possible, desirable, and inevitable (e.g., AI tutors). The dramatic acceleration in AI development following the launch of ChatGPT has been accompanied by a recent shift in global discourse, from an initial focus on safety and ethics to one that champions rapid action and innovation. This AI `gold rush' makes it more urgent than ever to develop a critical sensitivity to TSD. For AI to be developed responsibly and democratically, stakeholders must transition from passive reception of TSD's alluring narratives to active, critical engagement. This is the core rationale for this paper, which is based on the premise that for TSD to be effectively challenged, it must be systematically understood and identified.

The primary aim of this paper is to critically analyze the discourse of the AI executive elite to understand the presence, function, and dynamics of the Techno-Supremacy Doctrine. This research is guided by three objectives:
(I) to analyze whether and how aspects of TSD are presented, challenged, or reproduced; (II) to identify which discursive strategies are used to respond to AI-related concerns; and (III) to investigate how the elite's discourse has changed over time and provide potential explanations in terms of its evolving broader socio-political context.

This paper offers theoretical, methodological, and empirical contributions.

Theoretically, this paper articulates TSD as a comprehensive, synthesized theoretical framework that builds on established belief systems. It offers a structured way to understand how technical framing, solutionist impulses, chauvinistic beliefs, and quasi-religious faith can interlink to form a resilient, robust, and self-reinforcing worldview. Moreover, the concept of \emph{Benign Techno-Optimism} is introduced as a particular incarnation of TSD. This concept provides a crucial nuance to the study of techno-optimist discourse, moving beyond simplistic dichotomies (e.g., techno-optimism versus techno-pessimism). It allows for a more fine-grained analysis that can differentiate between uncritical, pernicious forms of TSD and more self-aware, critically tempered expressions of techno-optimism, providing a richer understanding of the spectrum of beliefs within the elite. 

Methodologically, in a field often characterized by high-level commentary, this paper develops a two-phase `T' approach to Critical Discourse Analysis (CDA) that systematically balances the need for a broad, corpus-wide overview with the necessity of in-depth qualitative analysis of specific phenomena. Novel metrics are introduced as quantitative heuristics. By translating detailed qualitative codings into transparent, comparable scores, these metrics provide a systematic way to identify patterns, variations, and outliers, thereby grounding the selection of texts for deeper analysis in verifiable data. This adaptation enhances the rigor and transparency of the analytical process, directly responding to scholarly critiques of CDA regarding potential analyst bias and a lack of systematic procedure.

Empirically, this paper employs CDA approach mentioned previously to analyze the public discourse of a systematically defined elite between 2017 and 2025. The analysis offers concrete, linguistically-grounded evidence of the spectrum of TSD adherence within this highly influential group, challenging monolithic portrayals of ``Silicon Valley thinking.'' The findings reveal the dynamic, adaptive nature of this elite discourse, including a marked polarization of stances in the post-ChatGPT era and significant shifts within the narratives of individual leaders. By tracing these dynamics and linking them to a detailed timeline of technological, political, and social events, this paper provides a crucial, evidence-based account of how the struggle to define the future of AI is being waged discursively by some of its most powerful proponents. 

After presenting TSD as the study’s theoretical foundation, the paper details the `T' approach to CDA developed for this research. It then describes the empirical study by first outlining the criteria for identifying the AI executive elite and selecting the research corpus, and then presenting findings in terms of TSD’s discursive spectrum and temporal dynamics. Key results and important considerations are then discussed, followed by a conclusion that highlights limitations and avenues for future research.

\section{The Techno-Supremacy Doctrine}
\label{sec:tsd}

\begin{quote}
``\textit{We believe that there is no material problem—whether created by nature or by technology—that cannot be solved with more technology}.'' - Marc Andreessen.
\end{quote}

The Techno-Supremacy Doctrine is a belief system characterized by excessive, often uncritical trust in technology's alleged inherent superiority to address complex societal challenges. TSD represents a mindset that overvalues technology's capacity to solve problems and often underestimates the complexity of social issues and the importance of human agency, values, and context. It fosters the idea that technological advancements alone are sufficient for societal progress, potentially leading to the neglect of non-technological solutions and a blindness to the limitations and potential negative consequences of purely technological approaches.

The term `Techno-Supremacy Doctrine' was selected over several alternatives for its precision in capturing this belief system's distinct ideological nature. The prefix `Techno' grounds the concept firmly in the domain of technology in its broadest sense, encompassing not just AI but technology in general. The core component, `Supremacy,' denotes more than mere primacy. It highlights a conviction in technology's inherent dominance, ultimate authority, and fundamental superiority over alternative non-technological approaches to problem-solving. This term reflects the overvaluation and uncritical trust central to TSD, where technological considerations often override or marginalize other vital human and contextual factors. Finally, `Doctrine' is employed because TSD represents not an isolated bias but a systemic and self-reinforcing set of interconnected beliefs, assumptions, and guiding tenets that shape perception and action. Note, however, that we do not claim that its tenets are always consciously held by its adherents.

TSD comprises four core components:
\begin{itemize}
    \item \emph{Tech Goggles} \citep{green_smart_2019}: Framing a wide array of societal issues primarily as technology problems (e.g., reducing fairness to a quantitative metric). Given a problem, Tech Goggles adherents tend to focus on how technology can address it.
    \item \emph{Solutionism} \citep{morozov_save_2013}: Regarding complex social, political, and human issues as problems with singular, often technological, solutions or ``fixes'' (e.g., believing education disparity can be primarily addressed with AI tutors). Solutionism often drives the rapid, widespread implementation of `solutions.'
    \item \emph{Techno-Chauvinism} \citep{broussard_artificial_2018}: Considering technology as inherently superior or objective. Techno-Chauvinists are firmly convinced that technology is the optimal solution to any problem, including complex societal issues, regardless of context (i.e., believing there is no problem technology cannot fix). 
    \item \emph{Cult of Technology} \citep{toyama_geek_2015}: Having a quasi-religious confidence in technology's alleged inherent power. Cult of Technology adherents are often overly optimistic about the future based on assumed, almost magical, capabilities of current or future technologies.
\end{itemize}

\begin{figure}
    \centering
    \includegraphics[width=0.5\linewidth]{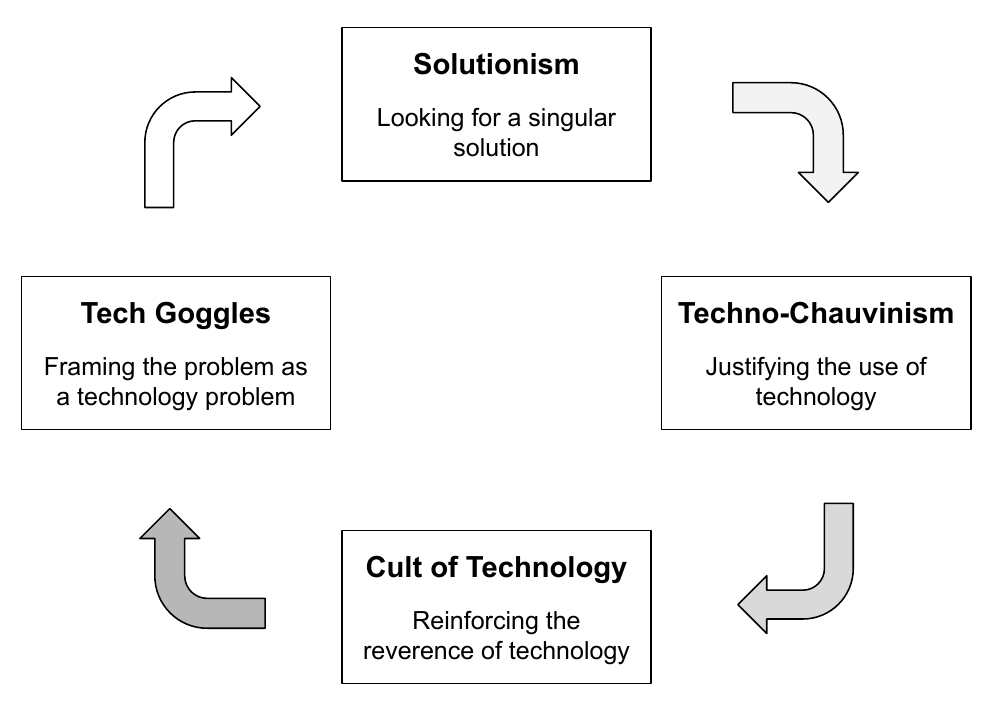}
    \caption{TSD Cycle}
    \label{fig:tsd_cycle}
\end{figure}

TSD's components contribute to a cycle of mutual reinforcement and amplification (cf. Figure \ref{fig:tsd_cycle}). Tech Goggles serves as an entry point, predisposing one to perceive and frame complex societal issues primarily through a technological lens. This framing naturally leads to Solutionism, narrowing the focus onto identifying singular technological ``solutions'' or ``fixes.'' Techno-Chauvinism provides justification, asserting the inherent superiority of the chosen approach over non-technological alternatives. The perceived effectiveness of this solution then fuels the Cult of Technology, strengthening the underlying faith in technology's power. This reinforced faith, in turn, makes the individual more inclined to habitually adopt Tech Goggles when facing future challenges, thereby perpetuating the cycle. 

\subsection{Related Concepts}

At its foundation, TSD is based on the premise of \emph{Technological Determinism} \citep{dotson_technological_2015, heder_ai_2021}, the belief that technology, often portrayed as an inevitable and unstoppable force, is the primary driver of social change (e.g., attributing the spread of literacy primarily to the invention of the printing press). TSD expands on this deterministic view, adding a normative dimension: technology-driven change is inherently progressive and superior to all other problem-solving strategies. It elevates the methods of \emph{computational thinking} \citep{wing_computational_2006} into a worldview, positing this approach as the optimal way to address complex societal problems. While a form of techno-optimism, TSD goes beyond a mere positive outlook to a doctrine of inherent superiority, sometimes manifesting as the quasi-religious faith seen in the Cult of Technology component.

TSD's tenets are expressed through various political and economic ideologies. Historically, it was manifest in the \emph{Californian Ideology} \citep{barbrook_californian_1996}, which blends libertarian and neoliberal principles with a belief in technology as a means of personal liberation that could bypass inefficient political institutions. More contemporary expressions range from \emph{Effective Accelerationism} (e/acc) \citep{cheok_effective_2024} to \emph{Transhumanism} \citep{hopkins_transhumanism_2012}. The former transforms TSD's tenets into a moral imperative to accelerate technological change at all costs, while the latter views overcoming human biology as the ultimate problem to be fixed. TSD has a dual relationship with \emph{Longtermism} \citep{steele_longtermism_2022}, which can either motivate critiques based on existential risks or reinforce TSD by framing the pursuit of AGI (Artificial General Intelligence) as a necessity for humanity's long-term future.

It is common for techno-optimists to assert that technology is neither good nor bad, but a value-neutral tool (cf. \emph{Value-Neutrality Thesis} (VNT) \citep{miller_is_2021}). TSD neither upholds nor denies VNT, as it merely claims technology's inherent \emph{instrumental} (rather than moral) superiority. However, by framing technology as uniquely powerful and scalable, but ultimately value-neutral, TSD proponents can deflect responsibility (cf. ``misuse,'' ``bad actors'') and portray risks as merely problems to be managed by better (e.g., more secure) technology, thereby reinforcing core TSD components like Tech Goggles and Solutionism.

\subsection{Concerns}

Concerns regarding TSD adherence can be grouped into three main areas: oversimplification of problems, devaluation of alternative approaches, and uncritical acceptance of technology.

Problem oversimplification arises significantly from the interaction of Tech Goggles and Solutionism. As Green and Morozov critique, this combination leads to framing diverse societal challenges primarily as narrow problems amenable to singular technological solutions. This framing can lead to a rush to adopt such solutions rapidly and widely before fully understanding the issue's nuances or even properly defining them (i.e., reaching for the answer before the question has been fully asked). This perspective inherently conflicts with deep contextual understanding and holistic problem assessment before intervention. It often oversimplifies or ignores the historical, social, economic, and political dimensions of societal issues that non-technological ethical frameworks prioritize. 

Techno-Chauvinism strongly drives the devaluation of alternative approaches. As Broussard critiques, this mindset inherently considers technology superior and more objective, thereby devaluing non-technological approaches as inferior (e.g., sub-optimal, inefficient, subjective). This stance directly opposes calls for critical evaluation of whether technology is the most appropriate approach, thereby neglecting alternative methods that might be more effective or ethically sound within a given social and political context. Moreover, this chauvinism often devalues \emph{situated knowledges} \citep{haraway_1988} from users and other affected stakeholders during technology design (unlike approaches like Participatory Design \citep{schuler_participatory_1993}).

The Cult of Technology fuels the uncritical acceptance of technology. Toyama's critique highlights the manifestation of this acceptance as an almost religious conviction in technology's alleged inherent transformative power, based on an overestimation of current or future capabilities. This blind faith discourages skepticism and rigorous testing, which are central to responsible innovation, often leading to the marginalization of potential negative consequences and the adoption of solutions without sufficient scrutiny, simply because they are technologically advanced. This stance contrasts with acknowledging that technology's impact is contingent on human agency and socio-institutional contexts, rather than an intrinsic magical quality. 

An overarching concern about TSD as a whole is that its adoption by people holding social, political, or economic power (e.g., the AI executive elite) directly undermines efforts towards responsible innovation. Adherence to TSD by influential people can lead to prioritizing technology research over other critical areas. Investments in social programs, policy reforms, ethical oversight, and human capital can be neglected, thereby hindering a holistic, ethical approach to societal progress.

While these critiques outline the potential risks associated with various aspects of TSD, they raise the question of whether more responsible and critically engaged manifestations could exist. A central goal of this paper's empirical analysis is to investigate the full spectrum of TSD in the analyzed discourse to determine if such nuanced positions exist and, if so, how they are linguistically realized.

\section{The `T' Approach to Critical Discourse Analysis}
\label{sec:cda}

This paper adopts and adapts Norman Fairclough's approach to Critical Discourse Analysis (CDA) \citep{fairclough_language_2015}. His framework rests on the fundamental assumption that \emph{discourse}, understood as language as a form of social practice, is not only influenced by broader social structures (e.g., the media, economic systems, the state) but also contributes to constituting, reproducing, and changing these structures. His approach consists of analyzing (written or spoken) text in three stages. The \emph{description} stage involves identifying the values embodied in the text's formal linguistic properties (e.g., vocabulary, grammar). The \emph{interpretation} stage aims to understand how the text was produced and could be interpreted to create meaning. Finally, the \emph{explanation} stage focuses on why discourse is the way it is and what its effects might be on the broader social context, particularly on power relations, ideologies, and social struggles.

CDA has been critiqued primarily around three areas: potential bias given the analyst's political positioning; lack of methodological rigor; and lack of empirical validation (cf.~\citep{ohalloran_critical_2022, stubbs_whorfs_1997, toolan_what_1997, widdowson_text_2008}). The methodology developed for this paper is specifically designed to address these concerns. The overall goal was to enhance analytical rigor and transparency while retaining the critical power of Fairclough's approach. To mitigate the risk of analyst bias, this research design combines high transparency (e.g., in the selection of authors and texts, in the annotation of texts) with critical self-reflexivity. Each text was consciously approached from positions of ``engagement and estrangement'' \citep{janks_critical_1997}, aiming to balance empathetic understanding with critical scrutiny. In response to critiques of methodological rigor, a two-phase `T' analysis was developed that includes deep qualitative investigations guided by patterns and hypotheses generated from a comprehensive coding of the entire corpus. The coding schemes developed provide a clear and consistent framework for identifying discursive features, thereby increasing the transparency and potential reproducibility of the analysis and mitigating the risk of cherry-picking data to fit a preconceived argument. Novel quantitative metrics were developed, derived directly from the coded linguistic data. This quantitative approach addresses critiques of CDA's empirical grounding by providing a concrete, data-driven basis for identifying patterns, comparing texts, and substantiating claims about the overall leanings and balance of the discourse. Finally, this paper adopts a focused scope for the explanation stage to address the critique concerning the empirical validation of the text-context link. The societal effects of discourse and audience cognitive reception are notoriously difficult to establish without extensive empirical methods, which are beyond the scope of this project. Therefore, the analysis refrains from making speculative claims about such impacts. Instead, the focus is on providing an explanatory analysis of potential \emph{causes}, grounding explanations in an examination of the socio-historical context.

\subsection{Broad Analysis}

\begingroup
\renewcommand{\arraystretch}{1.21}
\begin{table}[ht]
    \centering
    \begin{tabular}{p{9em} p{4em} p{12em} p{17em}} 
        \textbf{TSD Component}  & \textbf{Code} & \textbf{Name}                & \textbf{Description} \\
        \hline
        Tech Goggles            & TG-PE         & Prioritization of Efficiency & An overemphasis on efficiency and convenience as the primary values driving technology deployment and adoption. \\
        \cline{2-4}
                                & TG-SO         & Superior Outcomes            & A belief that addressing the technologically framed version of a problem effectively achieves the optimal or superior outcome for the underlying complex societal issue. \\
        \cline{2-4}
                                & TG-TF         & Technology Framing           & A tendency to perceive and define societal issues primarily through a technological lens. \\
        \hline
        Solutionism             & SL-SF         & Solution Focus               & A preference for narrow solutions to complex problems. \\
        \cline{2-4}
                                & SL-SP         & Singular Problem             & A tendency to assume a singular definition of a problem rather than thoroughly investigating its multifaceted nature. \\
        \cline{2-4}
                                & SL-UI         & Unquestioning Implementation & A drive to implement solutions rapidly and widely without sufficient critical examination of their potential consequences or underlying assumptions. \\
        \hline
        Techno-Chauvinism       & TC-AA         & Always the Answer            & An excessive conviction that technology, regardless of the context, is the optimal or primary solution to any problem. \\
        \cline{2-4}
                                & TC-OS         & Objectivity and Superiority  & A belief in the inherent objectivity and superiority of technology over non-technological approaches. \\
        \cline{2-4}
                                & TC-PD         & Progress Driver              & A tendency to frame technology as the essential, primary, or inevitable engine driving societal progress, advancement, and positive transformation. \\
        \hline
        Cult of Technology      & CT-DE         & Devaluation of Non-Tech      & A tendency to undervalue or disregard non-technological solutions, human wisdom, and social context in favor of technological interventions. \\
        \cline{2-4}
                                & CT-MP         & Magical Power                & A belief in the near-magical ability of technology to solve complex societal problems. \\
        \cline{2-4}
                                & CT-UF         & Utopian Future               & A belief in the transformative power of current or future technologies to create a better, even utopian world. \\
        \hline \\
    \end{tabular}
    \begin{minipage}{\linewidth}
        \textit{Note}. Analogous `anti' codes are used to annotate explicit contradictions of the corresponding aspect (e.g., ANTI-TG-PE). \\
    \end{minipage}
    \caption{Coding Scheme for TSD Presence and Reproduction}
    \label{tab:1_codes}
\end{table}
\endgroup

\begingroup
\renewcommand{\arraystretch}{1.21}
\begin{table}[ht]
    \centering
    \begin{tabular}{p{9em} p{4em} p{12em} p{17em}} 
        \textbf{Response Type}  & \textbf{Code} & \textbf{Name}     & \textbf{Description} \\
        \hline
        Acknowledging           & ACK-CR        & External Critique & Acknowledging a concern from others. \\
        \cline{2-4}
                                & ACK-RI        & Risk              & Acknowledging a concern. \\
        \hline
        Addressing              & ADD-JU        & Justification     & Addressing a concern, arguing that while possibly real, it is acceptable due to overriding benefits, necessity for progress, historical precedent, or context. \\
        \cline{2-4}
                                & ADD-RE        & Refutation        & Addressing a concern via a counter-argument against its validity. \\
        \cline{2-4}
                                & ADD-SN        & Non-Tech Solution & Addressing a concern via a proposed solution or mitigation that is \emph{not} primarily based on technology. \\
        \cline{2-4}
                                & ADD-ST        & Tech Solution     & Addressing a concern via a proposed solution or mitigation that is primarily based on technology. \\
        \hline
        Marginalizing           & MAR-DE        & Deflection        & Avoiding addressing a concern directly by changing the subject, shifting blame, or attacking the critic. \\
        \cline{2-4}
                                & MAR-DI        & Dismissal         & Rejecting a concern as fundamentally based on ignorance or fear-mongering, without substantive refutation. \\
        \cline{2-4}
                                & MAR-MI        & Minimization      & Treating a concern as minor, insignificant, rare, or exaggerated. \\
        \cline{2-4}
                                & MAR-RF        & Reframing         & Presenting a concern not as a problem but as neutral or beneficial. \\
        \hline \\
    \end{tabular}
    \caption{Coding Scheme for Response to AI-Related Concerns}
    \label{tab:2_codes}
\end{table}
\endgroup

\begingroup
\renewcommand{\arraystretch}{1.21}
\begin{table}[ht]
    \centering
    \begin{tabular}{lllc} 
        \textbf{(Sub-) Components}  &                                   & \textbf{Codes}                        & \textbf{Weight} \\ 
        \hline
        PRO-TSD                     & TSD Core Expression (TCE)         & CT-DE (Devaluation of Non-Tech)       & 2 \\
                                    &                                   & CT-MP (Magical Power)                 & \\
                                    &                                   & CT-UF (Utopian Future)                & \\
                                    &                                   & SL-SP (Singular Problem)              & \\
                                    &                                   & TC-AA (Always the Answer)             & \\
                                    &                                   & TC-PD (Progress Driver)               & \\
        \cline{3-4}
                                    &                                   & SL-SF (Solution Focus)                & 1 \\
                                    &                                   & SL-UI (Unquestioning Implementation)  & \\
                                    &                                   & TC-OS (Objectivity and Superiority)   & \\
                                    &                                   & TG-PE (Prioritization of Efficiency)  & \\
                                    &                                   & TG-SO (Superior Outcomes)             & \\
                                    &                                   & TG-TF (Technology Framing)            & \\
        \cline{2-4}
                                    & TSD-Reinforcing Responses (TRR)   & ADD-ST (Tech Solution)                & 2 \\
                                    &                                   & MAR-DI (Dismissal)                    & \\
                                    &                                   & MAR-MI (Minimization)                 & \\
        \cline{3-4}
                                    &                                   & ACK-CR (External Critique)            & 1 \\
                                    &                                   & ADD-JU (Justification)                & \\
                                    &                                   & ADD-RE (Refutation)                   & \\
                                    &                                   & MAR-DE (Deflection)                   & \\
                                    &                                   & MAR-RF (Reframing)                    & \\
        \hline
        ANTI-TSD                    & ANTI-TCE                          & \textit{Anti TCE codes with weight 2} & 2 \\
        \cline{3-4}
                                    &                                   & \textit{Anti TCE codes with weight 1} & 1 \\
        \cline{2-4}
                                    & ANTI-TRR                          & ADD-SN (Non-Tech Solution)            & 2 \\
        \cline{3-4}
                                    &                                   & ACK-RI (Risk)                         & 1 \\
        \hline \\
    \end{tabular}
    \begin{minipage}{\linewidth}
        \textit{Note}. The assignment of weights is a deliberate methodological choice grounded in the CDA principle of ideological centrality and discursive force. Specifically, greater weight is assigned to codes representing either core ideological tenets that form TSD's foundation or forceful discursive strategies that directly confront or dismiss opposing views. Conversely, a lesser weight is given to statements that function as more specific applications, manifestations, or indirect strategies.\\
    \end{minipage}
    \caption{Pro- and Anti-TSD Metric Components}
    \label{tab:metrics_components}
\end{table}
\endgroup

The first phase, represented by the top bar of the `T,' functions as an initial screening. It comprises a broad analysis across the corpus using relatively shallow versions of the CDA stages of description and interpretation, followed by a quantitative analysis to identify patterns for investigation in the second phase. The goal of this phase is to gain a detailed understanding of each author's adherence to TSD (cf. objective I) and their responses to AI-related concerns (cf. objective II).

This phase is composed of three steps. First, upon an initial reading, each text's \emph{point} (i.e., the summary interpretation of the text intended by the author) and its likely intended target audience (e.g., peers, potential followers, adversaries) are identified. Second, on a closer reading, individual sentences are annotated with custom codes corresponding to objectives I (cf. Table \ref{tab:1_codes}) and II (cf. Table \ref{tab:2_codes}), respectively. For example, the sentence ``Technology will usher in an era of prosperity for all mankind.'' would be annotated with codes TC-PD (Progress Driver) and CT-UF (Utopian Future). Finally, a quantitative analysis of code counts is conducted to determine the extent to which each text is pro- or anti-TSD. Each text is analyzed both in isolation and with respect to other texts to identify concrete phenomena to be qualitatively investigated in the second phase of the analysis.

Two metrics were developed for the quantitative analysis: \emph{TSD Adherence} (TDSA) and \emph{TSD Balance} (TSDB). TSDA aims to synthesize the prevalence of pro- and anti-TSD discourse and how responses to AI-related concerns might reinforce or counteract TSD within each analyzed text. Conversely, TSDB measures the balance between pro- and anti-TSD discourse within each text. Used together, these metrics enable a rich classification, allowing for a distinction between one-sided advocacy (high positive TSDA, low TSDB), one-sided critique (high negative TSDA, low TSDB), and texts characterized by significant internal debate (TSDA near zero, high TSDB). These metrics were developed as heuristics to identify broad discursive patterns and should not be considered absolute or objective measures of an individual's beliefs. Their function is merely to guide and ground deeper qualitative analyses by providing comparative data across the corpus.

TSDA and TSDB are defined in terms of pro- and anti-TSD components, denoted PRO-TSD and ANTI-TSD, respectively. TSDA is defined as their difference, while TSDB is defined as the percentage of the smaller with respect to the sum of both:
\begin{equation*}
\begin{split}
    \text{TSDA} & = \text{PRO-TSD} - \text{ANTI-TSD} \\
    \text{TSDB} & = \frac{\min(\text{PRO-TSD}, \text{ANTI-TSD})}{\text{PRO-TSD} + \text{ANTI-TSD}}
\end{split}
\end{equation*}
PRO-TSD and ANTI-TSD are defined as a weighted sum of the normalized counts of specific codes according to Table \ref{tab:metrics_components}.\footnote{Counts are normalized to ensure text comparability by dividing raw counts by the text's number of words and multiplying by 1,000.} For example, suppose a particular 2,000-word text has 6 sentences annotated with TC-PD and 4 with TG-PE. By normalizing counts and multiplying by the corresponding weights, we have that $\text{TCE} = ((6 / 2000 * 1000) * 2) + ((4 / 2000 * 1000) * 1) = 8$. Suppose that TRR = 2, ANTI-TCE = 1, and ANTI-TRR = 4. Then, $\text{PRO-TSD} = \text{TCE} + \text{TRR} = 10$ and $\text{ANTI-TSD} = \text{ANTI-TCE} + \text{ANTI-TRR} = 5$. Therefore, $\text{TSDA} = 5$ and $\text{TSDB} = 0.\overline{3}$. According to these values, this text is pro-TSD ($\text{TSDA} > 0$) and presents its stance in a moderately balanced way ($\text{TSDB} > 0.25$).\footnote{Perfect balance ($\text{TSDB} = 0.5$) is achieved if both components are the same (but different than 0), while perfect one-sidedness ($\text{TSDB} = 0$) is achieved if only one is 0. If both are 0, then TSDB is undefined, but this can only happen for irrelevant texts with no annotations.}

\subsection{Deep Analysis}
The second phase of the methodology, represented by the vertical stem of the `T,' comprises deeper investigations of selected phenomena to understand their functions and manifestations in detail. This phase aims to inspect concrete phenomena identified in the previous phase more deeply and in context through a qualitative lens, thereby complementing the previous, mainly quantitative analysis. It involves applying the full-fledged CDA stages of description, interpretation, and/or explanation.

The description stage involves analyzing linguistic features for the values they embody: \emph{experiential}, \emph{expressive}, and/or \emph{relational}. Experiential values are related to how the author represents the world (i.e., how the author frames problems and envisions potential solutions). These values are mainly tied to Tech Goggles and Solutionism. Expressive values are related to the author's evaluations and stances (i.e., what problem-solving approaches the author values and prefers). These values are mainly tied to Techno-Chauvinism and the Cult of Technology. Finally, relational values correspond to the relationships the author assumes, establishes, or negotiates with the intended audience. The interpretation stage involves analyzing how the linguistic features identified during the description stage function communicatively to establish the text's point and purpose. This stage requires analyzing each text's \emph{situational} and \emph{intertextual} contexts (i.e., its production and interpretation conditions, and the historical series to which it belongs). Finally, the explanation stage involves analyzing discursive strategies over time in the context of broader power struggles. The goal is to suggest potential explanations in terms of \emph{social determinants} for why the discourse has changed in the way it has (cf. objective III).\footnote{This stage also involves analyzing the discourse in terms of its social effects; however, due to time, space, and methodological constraints, this paper focuses solely on social determinants.} Though presented sequentially, these stages are intended to be iterative. Fairclough suggests that the analyst move between them to build a holistic, mutually explanatory account.

\section{Empirical Study}
\label{sec:emp}

The empirical study consisted of applying the `T' CDA methodology described previously to a corpus of publicly available texts authored by members of the AI executive elite. This elite comprises individuals with significant decision-making authority over the way AI is funded, developed, and deployed globally. It is defined by four criteria. First, the CEOs and CTOs of the top five companies by market share in generative AI models. As of writing, these are Microsoft (including Microsoft AI) (39\%), Amazon (including AWS) (19\%), Google (including Google Deepmind) (15\%), OpenAI (9\%), and Anthropic (4\%) \citep{fernandez_leading_2025}. Second, the Managing Partners (MPs) of the top ten venture capital firms by assets under management, provided they have invested in the aforementioned companies exclusively devoted to AI (i.e., OpenAI and Anthropic). As of writing, these are Andreessen Horowitz, General Catalyst Partners, Sequoia Capital, and Khosla Ventures \citep{noauthor_anthropic_2025, noauthor_top_2025, peck_top_2025}. Third, the CEO and CTO of the leading provider of Graphics Processing Units, Nvidia \citep{fernandez_leading_2025}. Finally, two individuals, Mark Zuckerberg and Elon Musk, due to their influential public profiles and substantial involvement in the AI field. Elite members were identified in March 2025 via Google Search, with all data rigorously verified against the corresponding official company and firm websites. The final group comprises 19 individuals: 11 CEOs, three CTOs, and five MPs (cf. Table \ref{tab:elite}).

\begingroup
\renewcommand{\arraystretch}{1.21}
\begin{table}[ht]
    \centering
    \begin{tabular}{lll}
        \textbf{Name}       & \textbf{Role} & \textbf{Company} \\
        \hline
        Andy Jassy          & CEO           & Amazon \\
        Werner Vogels       & CTO           & Amazon \\
        Dario Amodei        & CEO           & Anthropic \\
        Matt Garman         & CEO           & AWS \\
        Sundar Pichai       & CEO           & Google \\
        Demis Hassabis      & CEO           & Google DeepMind \\
        Satya Nadella       & CEO           & Microsoft \\
        Kevin Scott         & CTO           & Microsoft \\
        Mustafa Suleyman    & CEO           & Microsoft AI \\ 
        Sam Altman          & CEO           & OpenAI \\
        \hline
        Marc Andreessen     & MP            & Andreessen Horowitz \\
        Ben Horowitz        & MP            & Andreessen Horowitz \\
        Hemant Taneja       & MP            & General Catalyst Partners \\
        Vinod Khosla        & MP            & Khosla Ventures \\
        Roelof Botha        & MP            & Sequoia Capital \\
        \hline
        Jensen Huang        & CEO           & Nvidia \\
        Michael Kagan       & CTO           & Nvidia \\
        \hline 
        Mark Zuckerberg     & CEO           & Meta \\
        Elon Musk           & CEO           & xAI \\
        \hline \\
    \end{tabular}
    \caption{AI Executive Elite}
    \label{tab:elite}
    \begin{minipage}{\linewidth}
        \textit{Note}. While other notable individuals (e.g., chief scientists, prominent academic researchers, influential journalists) unquestionably contribute to shaping the AI discourse, by focusing specifically on C-suite executives and MPs, this paper intentionally prioritizes the analysis of ideological discourse linked to direct developmental and financial power. The cultural and scientific influence exerted by other actors is critical, but it represents a different form of power beyond the scope of this investigation.\\
    \end{minipage}
\end{table}
\endgroup

\begingroup
\renewcommand{\arraystretch}{1.21}
\begin{table}[ht]
    \centering
    \begin{tabular}{ll}
        \textbf{Title}                                                                          & \textbf{Identifier} \\
        \hline
        \textit{Who will control the future of AI?} \citep{altman_who_2024}                     & \texttt{altman-2024-control} \\
        \textit{The Intelligence Age} \citep{altman_intelligence_2024}                          & \texttt{altman-2024-intelligence} \\
        \textit{Remarks on Anthropic's Responsible Scaling Policy} \citep{amodei_dario_2023}    & \texttt{amodei-2023-scaling} \\
        \textit{Machines of Loving Grace} \citep{amodei_machines_2024}                          & \texttt{amodei-2024-grace} \\
        \textit{Why AI Will Save the World} \citep{andreessen_why_2023}                         & \texttt{andreessen-2023-save} \\
        \textit{The Techno-Optimist Manifesto} \citep{andreessen_techno-optimist_2023}          & \texttt{andreessen-2023-manifesto} \\
        \textit{AI: Scary for the Right Reasons} \citep{khosla_ai_2017}                         & \texttt{khosla-2017-scary} \\
        \textit{AI: Dystopia or Utopia?} \citep{khosla_ai_2024}                                 & \texttt{khosla-2024-utopia} \\
        \textit{Why Google thinks we need to regulate AI} \citep{pichai_why_2020}               & \texttt{pichai-2020-regulate} \\
        \textit{The AI Action Summit: A golden age of innovation} \citep{pichai_ai_2025}        & \texttt{pichai-2025-innovation} \\
        \textit{The Technology of Intelligence} \citep[c.~4]{suleyman_coming_2023}              & \texttt{suleyman-2023-intelligence} \\
        \textit{Containment for AI} \citep{suleyman_containment_2024}                           & \texttt{suleyman-2024-containment} \\
        \textit{The AI Century} \citep[c.~2]{taneja_unscaled_2018}                              & \texttt{taneja-2018-century} \\
        \textit{The End of Unintended Consequences} \citep[c.~1]{taneja_intended_2022}          & \texttt{taneja-2022-consequences} \\
        \hline \\
    \end{tabular}
    \caption{Corpus\label{tab:corpus}}
\end{table}
\endgroup

The corpus contains texts in which elite members discuss the impact of technology or AI on society, including opportunities and concerns. It was constructed systematically in accordance with concrete criteria to avoid cherry-picking. Eligible texts are book chapters, op-eds, and blog posts (personal or corporate) of at least 750 words published since 2017. These criteria are intended to include texts in which members of the AI executive elite substantively express their views, allowing for in-depth CDA, which is why shorter texts (e.g., tweets) were omitted. The year 2017 was chosen because it marks a pivotal year for generative AI, with the seminal paper \emph{Attention Is All You Need} \citep{vaswani_attention_2017} laying the theoretical foundations for modern AI. Unless published as blog posts or op-eds, spoken texts were omitted as they tend to be less concise, more prone to vagueness, significantly longer, and more difficult to analyze due to non-verbal communication, all of which pose challenges given the time constraints of this research.

Texts were gathered in March 2025 using Google Search. The initial pool contained 32 texts. Given the need for in-depth analysis and time constraints, the final corpus was selected from two texts by authors with the most substantial engagement on the topic. Author engagement was measured in terms of the total number of words written ($w$) and the number of distinct publication dates ($d$) as $E = (w' + d')/2$, where $w'$ and $d'$ are the min-max normalization of $\log(w)$ and $d$, respectively.\footnote{This normalization is used to rescale values to a range between 0 and 1. The logarithmic transformation is applied to the number of words because the distribution is very skewed (i.e., some individuals have written only a few hundred words, while others have written several thousand).} Only authors with an engagement score higher than 0.5 were considered for this research. While all eligible authors wrote at least two texts in the initial pool, most wrote more. The final two texts per author were selected by prioritizing text type diversity (e.g., a blog post and an op-ed are preferred over two blog posts), topic specificity (i.e., AI over technology in general), and length (i.e., longer texts over shorter ones) in that order.

The final corpus is shown in Table \ref{tab:corpus}. Surprisingly, some influential and vocal individuals (e.g., Hassabis, Musk) have not published any eligible texts, suggesting that their ideological influence is projected through different media (e.g., social media posts, interviews, podcasts, product demos).

\subsection{The Spectrum of TSD Discourse}

\begin{quote}
``\textit{Our enemy is the Precautionary Principle, which would have prevented virtually all progress since man first harnessed fire.}'' - Marc Andreessen.    
\end{quote}

\begin{quote}
``\textit{The goal should be to create a set of interlinked and mutually reinforcing technical, cultural, legal, and political mechanisms for maintaining societal control of AI.}'' - Mustafa Suleyman.
\end{quote}

\begin{figure*}[ht]
    \centering
    \includegraphics[width=1\linewidth]{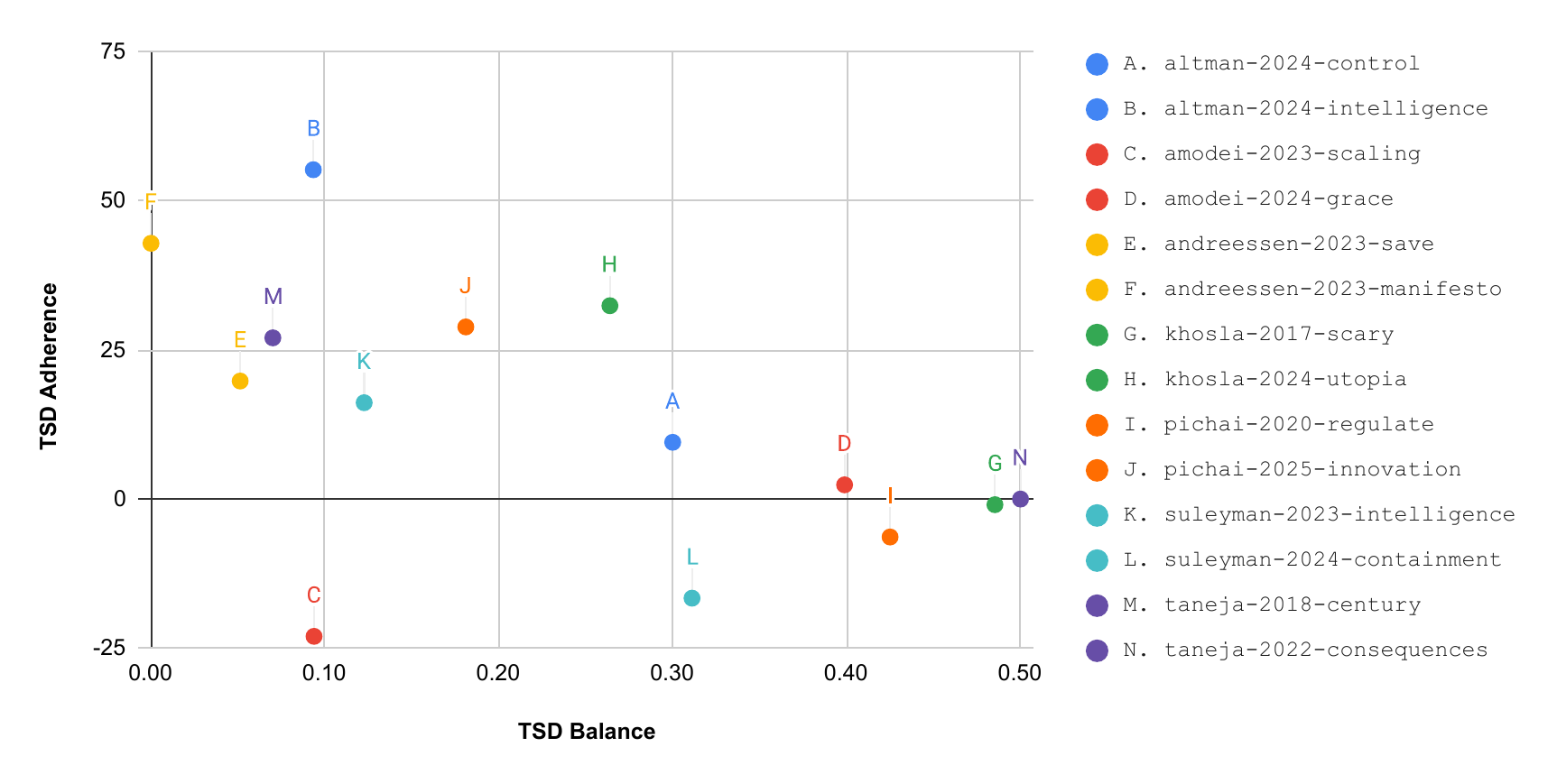}
    \caption{TSD Discourse Spectrum}
    \label{fig:tsd_spectrum}
\end{figure*}

Figure \ref{fig:tsd_spectrum} shows the stances' spectrum in terms of TSD Adherence and Balance. On the Y-axis (TSDA), higher positive values mean more pro-TSD leaning discourse, whereas negative values mean more anti-TSD leaning discourse. The X-axis (TSDB) ranges from 0, corresponding to completely one-sided discourse (either pro- or anti-TSD), to 0.5, representing perfectly balanced discourse. The top-left region (high TSDA, low TSDB) represents imbalanced, strongly pro-TSD texts. The bottom-left region (negative TSDA, low TSDB) represents imbalanced, strongly anti-TSD texts.\footnote{The region along the Y-axis where TSDB is high will necessarily contain texts with TSDA scores close to zero. This is because a TSDB value close to 0.5 indicates that the pro- and anti-TSD components of TSDA are very similar, leading to near or complete cancellation.}

The typical text in the corpus tends to be moderately pro-TSD, presenting this stance in a relatively imbalanced way. Notably, however, both TSDA and TSDB have very high standard deviations, indicating that the AI executive elite is not a monolithic group in how TSD is expressed and how AI-related concerns are addressed. While, on average, texts tend to be pro-TSD, some texts are strongly pro-TSD (e.g., \texttt{\small altman-2024-intelligence}), strongly anti-TSD (e.g., \texttt{\small amodei-2023-scaling}), or much more neutral (e.g., \texttt{\small khosla-2017-scary}). Similarly, while pro- or anti-TSD discourse is typically moderately more dominant within a given text, some texts are notably imbalanced (e.g., \texttt{\small andreessen-2023-manifesto}), while others are remarkably balanced (e.g., \texttt{\small taneja-2022-consequences}).

Texts cluster into three main profiles. The first profile (top-left to mid-left on the chart) is characterized by a strong pro-TSD stance presented in a fairly imbalanced way. Both of Andreessen's texts are extreme examples of this one-sided pro-TSD discourse. The second profile (bottom half of the chart) contains critical and counter-TSD texts. \texttt{\small amodei-2023-scaling} is an outlier, not only heavily anti-TSD in its overall message but also quite imbalanced in that direction. While \texttt{\small suleyman-2024-containment} also has anti-TSD leaning, it is achieved with more internal balance. Both \texttt{\small pichai-2020-regulate} and \texttt{\small khosla-2017-scary} exhibit a slight anti-TSD lean, achieved through a relatively balanced presentation of pro- and anti-TSD elements. The third profile (on the right side of the chart) contains balanced texts (e.g., \texttt{\small khosla-2017-scary}, \texttt{\small taneja-2022-consequences}).

While TSDA and TSDB provide valuable insights, it is important to approach them with nuance. For example, while \texttt{\small taneja-2022-consequences} displays an equilibrium between the pro- and anti-TSD components, this ``perfect balance'' does not render this text ideal. A high TSDB score indicates proportional discursive engagement with both sides as defined by the coding scheme. However, the qualitative nature and potential subtlety of specific TSD manifestations, some of which may be more insidious than others, require in-depth qualitative analysis.

Discourse reproducing TSD's core tenets is more voluminous than TSD-reinforcing responses to AI-related concerns. Conversely, anti-TSD responses to AI-related concerns are more prevalent than explicit contradictions of core TSD aspects, which are extremely rare. Two specific TSD core aspects are by far the most common across the corpus: (i) believing that technology will create a better, even utopian world and (II) framing technology as the primary or inevitable engine driving positive societal transformation. Conversely, the most prevalent response to concerns across the corpus is explicit acknowledgment; however, as will be discussed later, this is often used merely as a preamble to pro-TSD discourse.

The analysis reveals a distinct TSD manifestation, referred to here as \emph{Benign Techno-Optimism} (BTO). This type of discourse is characterized by a high degree of Cult of Technology expressions, particularly regarding utopian futures (CT-UF) and the almost magical power of technology (CT-MP). However, these optimistic expressions are critically tempered by significant counterarguments to TSD. These arguments include direct contradictions of TSD tenets, especially those challenging the idea of technology as an inevitable driver of progress (ANTI-TC-PD), as well as critical responses to AI-related concerns, particularly proposals for solutions that are not primarily based on technology (ADD-SN). Both \texttt{\small suleyman-2024-containment} and \texttt{\small amodei-2024-grace} exemplify BTO.

The remainder of the section presents a discussion on two discursive phenomena selected for their high frequency in the quantitative analysis and their central importance to the reproduction of TSD.

\subsubsection{AI-Driven Utopia}
Believing in an (almost) utopian future (CT-UF), and considering technology the main or inevitable driver of societal progress (TC-PD) are the most expressed core TSD aspects. Moreover, they are employed \emph{together} in ten of the 14 analyzed texts (71\%).

This discursive strategy is exemplified by \texttt{\small altman-2024-intelligence}. This text is likely intended to shape public opinion about the trajectory and future of AI. It engages with and builds upon previous narratives about technological progress and economic transformation, aiming to counter dystopian concerns with an overwhelmingly positive vision (e.g., \texttt{\small andreessen-2023-manifesto}, \texttt{\small khosla-2024-utopia}). Consider the following excerpt:

\begin{quote}
With these new [AI] abilities, we can have shared prosperity to an unimaginable degree today; in the future, everyone's lives can be better than anyone's life is now.

Here is one narrow way to look at human history: after thousands of years of compounding scientific discovery and technological progress, we have figured out how to melt sand, add some impurities, arrange it with astonishing precision at extraordinarily tiny scale into computer chips, run energy through it, and end up with systems capable of creating increasingly capable artificial intelligence.

This may turn out to be the most consequential fact about all of history so far. \citep[p.~2]{altman_intelligence_2024}    
\end{quote}

The first sentence presents an AI-driven utopian future, making the reader receptive to the idea that large-scale AI development is desirable. Its declarative nature positions the author as a visionary who understands AI's true, transformative potential. Agency is implicitly given to AI (``AI abilities''), while humanity (``we'') is portrayed as a passive beneficiary. The sentence highlights betterment (``prosperity,'' ``better''), its magnitude (``unimaginable,'' ``everyone's lives can be better than anyone's life''), and its universality (``we,'' ``shared,'' ``everyone''), preemptively countering criticisms that AI may exacerbate social inequalities. The juxtaposition of ``today'' and ``the future'' emphasizes AI's potential (rather than its shortcomings). The author employs \emph{rewording} (i.e., expressing the same idea in different ways) by connecting two somewhat redundant and reinforcing claims. Although the author uses modal verbs (``can have,'' ``can be'') to express potential and capability, the declarative mood of the sentence portrays these possibilities as inevitable or highly likely. The second sentence details a technological development process culminating in AI, which legitimizes it as the logical and natural pinnacle of thousands of years of human endeavor. Agency is explicitly given to humanity (``we have figured out''), but it is in service of technological development. The author's admittedly ``narrow'' portrayal of human history is exclusively focused on the role of science and technology (``compounding scientific discovery and technological progress''). The use of expressive words (``astonishing,'' ``extraordinarily'') highlights the author's marvel of technological achievement. The sentence is long and cumulative, depicting a linear progress path culminating in AI. The narrative around ``compounding'' progress creates a sense of inevitability and inherent momentum in technological development. Finally, the third sentence builds directly on the previous one, underscoring AI's unparalleled importance and implicitly justifying the immense resources directed towards it. Even though a modal verb (``may'') softens the claim, its superlative nature powerfully portrays technology as the ultimate driver of progress on a simplistic linear path.

This discursive strategy creates a compelling narrative. It first portrays AI as the inevitable culmination of technological progress, and then heralds an era of unparalleled, universally shared prosperity driven by that progress. This dual assertion of historical inevitability and utopian promise functions to legitimize current AI development and deployment paradigms, justifying immense investment and focus. Additionally, this narrative can be used to marginalize critical perspectives or calls for caution as ``anti-progress'' or unduly pessimistic (cf. ``AI doomers''). In doing so, it subtly downplays the intricate requirements of social adaptation, robust governance, and inclusive human agency essential for the responsible development of AI.

\subsubsection{Pro-TSD Risk Acknowledgment}
While explicit acknowledgment of risk is the most common anti-TSD response to concerns, the data reveal that it was never the final word in any of the analyzed texts. In 13 of the 14 texts (92\%), risk acknowledgment was followed by a discursive move aimed at managing, justifying, or resolving the concern in a pro-TSD manner (the single text where this did not occur is \texttt{\small andreessen-2023-manifesto} because it acknowledges no concerns at all).

This discursive strategy is exemplified by \texttt{\small taneja-2022-consequences}. This text is part of a book that aims to redefine how technology is developed and invested in, moving towards a more responsible model in which technology must be ethically sound and beneficial. It is part of a narrative ecosystem responding to growing public and regulatory concerns about the negative impacts of technology, offering forward-looking, constructive paths that align technological development with societal good (e.g., \texttt{\small pichai-2020-regulate}, \texttt{\small amodei-2023-scaling}, \texttt{\small suleyman-2024-containment}). Consider the following excerpt:

\begin{quote}
Throughout history, technology has often been great. But now it must also be good.

Companies that create technology that does good and avoids harmful, unintended consequences will win in the years to come. Investing for financial returns and for positive impact are no longer two different things. They are now the same thing.

Today's powerful technologies have too much destructive potential [...] Now, imagine how Moore's law and Metcalfe's law could accelerate the damage done by a biased or rogue artificial intelligence. \cite[p.~16]{taneja_intended_2022}
\end{quote}

The first paragraph establishes the core message: it is no longer adequate for technology to be merely ``great;''; it now needs to be ``good.'' This paragraph, by distinguishing between greatness and goodness, acknowledges that the previous state of affairs was lacking. The author presents technology in a positive light (``often been great''), but this is immediately followed by a crucial turning point (``But'') emphasized by the juxtaposition of the past (``Throughout history'') and the present (``now''). The use of the modal verb ``must'' expresses necessity. Both sentences are declarative and straightforward, conveying authority and conviction. The second paragraph claims that companies will be incentivized to ``create technology that does good,'' not only on moral grounds but also on financial and business grounds. This idea creates a business case for responsible innovation, where responsible companies can still ``win,'' countering narratives that portray social responsibility as fundamentally in tension with financial success. The alignment between moral responsibility and capitalist incentives is emphasized through the use of rewording (``are no longer two different things,'' ``They are now the same thing''). This alignment makes the author's argument palatable to a business and investment audience. Agency is explicitly given to companies (``Companies that create''), and a new criterion for successful technology is established (``technology that does good and avoids harmful, unintended consequences'') using the future tense (``will'') to convey conviction and authority. The third paragraph acknowledges that current technologies have ``too much destructive potential'' and highlights the potential harm that ``a biased or rogue'' AI could inflict. This paragraph validates the proposed strategy and portrays the shift from mere greatness to goodness as necessary and urgent. The imperative mood (``Imagine'') directly asks the reader to consider the severity of the risk. While the use of a modal verb (``could'') softens the certainty of the threat, mentioning that Moore's and Metcalfe's laws are ``accelerating damage'' provides concrete, technologically grounded reasons for the urgency.

The combination of risk acknowledgment with the advancement of a technology-based solution is particularly insidious. The initial acknowledgment validates the reader's concerns, but the immediate pivot to technological fixes implicitly frames these risks as ultimately problems that can be solved by further technological innovation. The insidious nature of this type of discourse lies in its power of \emph{naturalization}, a key mechanism of ideological power in Fairclough's framework. By framing the conversation as `risk acknowledged, tech solution proposed,' the discourse makes the pivot to a technological fix seem commonsensical and natural. When repeated over time, this pattern naturalizes the underlying assumption that the appropriate response to technological harm is always more or better technology, effectively marginalizing non-technological solutions until they seem impractical or irrelevant. The ideology becomes most potent when it is no longer perceived as ideology, but simply as the way things are naturally done. Truly anti-TSD discourse (e.g., BTO) would not only acknowledge risk but also question core TSD tenets and comprehensively explore diverse solutions, including non-tech proposals.

\subsection{The Dynamics of TSD Discourse}

\begin{quote}
``\textit{Good regulatory frameworks will consider safety, explainability, fairness and accountability to ensure we develop the right tools in the right ways.}'' - Sundar Pichai, 2020.    
\end{quote}

\begin{quote}
``\textit{Successful policy addresses risks, without stymying innovation, progress and the positive impacts.}'' - Sundar Pichai, 2025.
\end{quote}

\begin{figure*}[ht]
    \centering
    \includegraphics[width=1\linewidth]{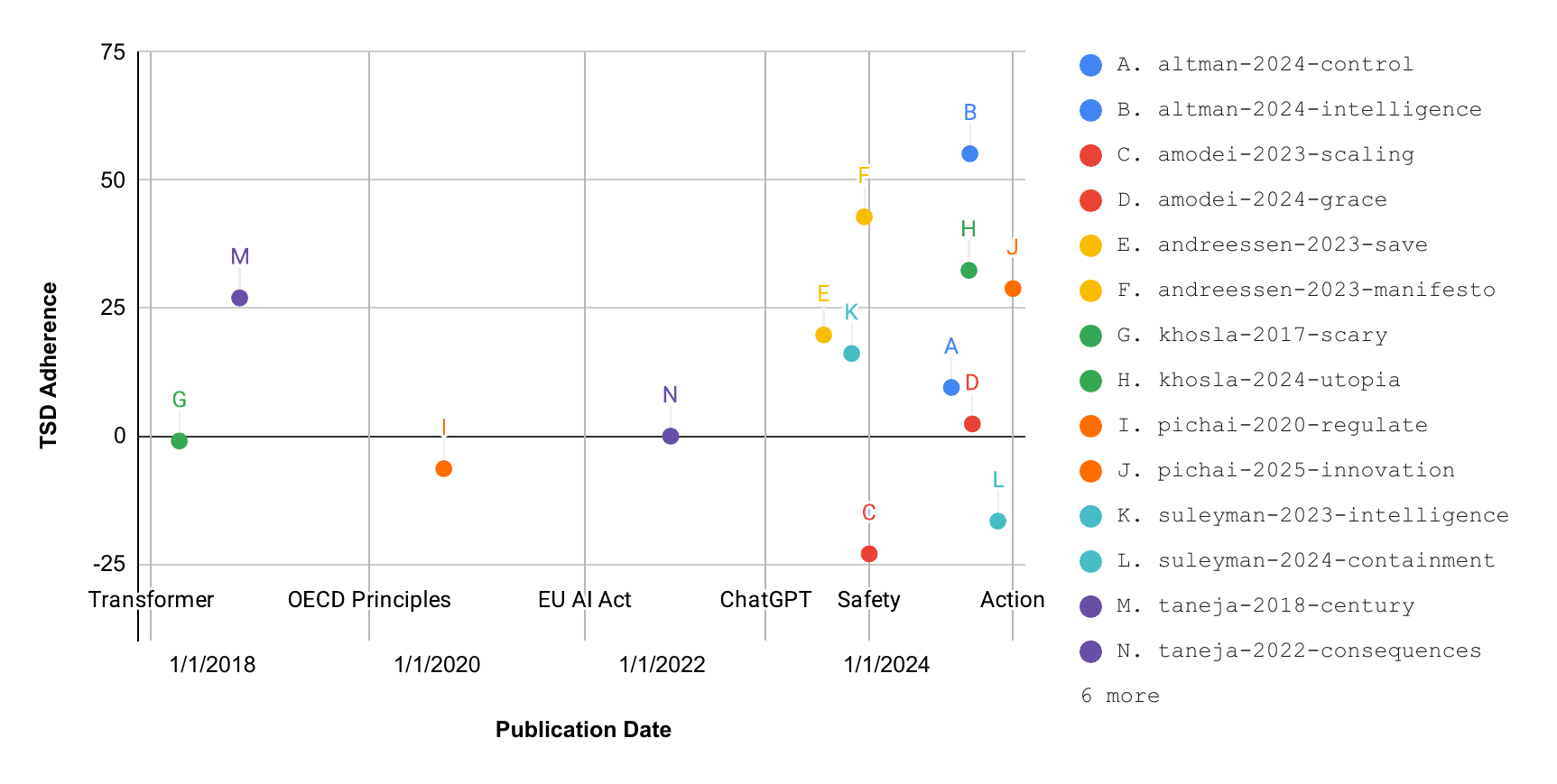}
    \caption{TSD Discourse Dynamics}
    \label{fig:tsd_dynamics}
\end{figure*}

Figure \ref{fig:tsd_dynamics} shows TSD Adherence over time. The chart is contextualized with key milestones in technological capability (e.g., the launch of ChatGPT), foundational ethical and policy framings (e.g., the EU AI Act), and international governance dialogues (e.g., the AI Safety Summit). More details on this broader context are provided in the following section, which serves as a backdrop for the subsequent analysis.

The data reveal distinct patterns of discursive change over time, with most authors demonstrating significant shifts toward more pronounced and imbalanced pro-TSD stances. For example, Pichai moved from a balanced, slightly anti-TSD position in 2020 to a strongly pro-TSD, imbalanced position in 2025. Similarly, Khosla shifted from a highly balanced, neutral stance in 2017 to a strongly pro-TSD, imbalanced stance in 2024. This dynamic can be rapid; within the same year, Altman moved from a moderately pro-TSD text to one that is extremely pro-TSD and imbalanced.

Beyond these individual trajectories, a broader trend of polarization emerges across the corpus, particularly in the period following the launch of ChatGPT. The period from late 2023 onwards exhibits the most extreme TSDA scores in both positive and negative directions. Both the highest pro-TSD scores observed in the corpus (\texttt{\small altman-2024-intelligence}, \texttt{\small andreessen-2023-manifesto}, \texttt{\small khosla-2024-utopia}, \texttt{\small pichai-2025-innovation}) and the most significantly anti-TSD scores (\texttt{\small amodei-2023-scaling}, \texttt{\small suleyman-2024-containment}) are all concentrated in this recent period. Overall, five of the seven authors (71\%) showed an increase in their TSDA scores over time, with four of these five (80\%) also becoming less balanced. Suleyman is the only author whose TSDA score decreases (and TSDB score increases) after the launch of ChatGPT.

As cautioned in the previous section, it is crucial to interpret these figures with nuance. Specifically, differences in scores between texts by the same author, particularly those published relatively close in time, do not necessarily indicate a trend or a fundamental shift in their convictions. Rather, such variations may reflect strategic discursive adaptations tailored to different communicative contexts. The purpose of the text, the intended audience, the specific forum or publication venue, and the immediate events surrounding the text's release can all significantly influence the discursive strategies employed. In-depth qualitative analysis is necessary to gain a deeper understanding of the nature and potential motivations behind such discursive differences, moving beyond purely linear, evolutionary interpretations.

The remainder of the section provides a qualitative analysis of the recent surge of pro-TSD discourse. The broader societal context in which relevant texts were produced is first described. Then, a discussion on potential social determinants contributing to this discursive shift is offered.

\subsubsection{From Safety to Action}
The period following ChatGPT's launch in late 2022 has witnessed a dramatic acceleration in AI development and a shift in the dominant discourse among the elite. While the initial generative AI boom was characterized by a robust debate on long-term safety and ethics, this was soon challenged and eclipsed by a powerful counter-narrative emphasizing rapid innovation and geopolitical competition. The new American administration epitomized this shift at the global AI Action Summit in early 2025, where Vice President James David Vance explicitly declared that he was there to discuss opportunity, not safety \citep{noauthor_remarks_2025}.

As AI capabilities surged after the publication of the Transformer architecture in 2017, the demand for ethical guidelines and regulatory frameworks intensified. Building on a decade of several high-level and voluntary ethics principles from corporations (e.g., Google's \citep{pichai_ai_2018}) and organizations (e.g., OECD's \citep{noauthor_ai-principles_2019}), the European Union AI Act, first proposed in 2021, signaled a concrete move towards binding regulation \citep{noauthor_eu_2023}. The unprecedented user adoption of ChatGPT put AI into the global public consciousness. Throughout 2023, media coverage increasingly fixated on concerns such as algorithmic bias, the dangers of misinformation, and AI-driven job displacement. These anxieties were amplified by seminal figures like Geoffrey Hinton, who resigned from Google to warn of long-term risks, and by the Future of Life Institute's open letter, urging to ``immediately pause the training of AI systems more powerful than GPT-4'' in March 2023 \citep{noauthor_pause_2023}. In October 2023, then US President Joseph Biden issued a comprehensive Executive Order \citep{noauthor_safe_2023}, focusing on AI safety ahead of the first global AI Safety Summit hosted by the United Kingdom in November. The summit culminated in the Bletchley Declaration, in which 28 countries, including the US and China, acknowledged the potential for ``serious, even catastrophic, harm'' and called for international cooperation \citep{noauthor_bletchley_2023}. The subsequent AI Seoul Summit in May 2024 saw 16 leading AI companies make AI safety commitments; however, it subtly foreshadowed the changing consensus by explicitly adding innovation to the global agenda.

A confluence of powerful technological, commercial, and political forces fueled the discursive shift towards action and innovation. Technologically, the past several years have witnessed an explosion of breakthroughs, including the development of increasingly powerful hardware (e.g., Nvidia's GPU architecture series), models (e.g., OpenAI's GPT series; Google's BERT, Bard, LaMDA, and Gemini series; Anthropic's Claude series), and practical applications (e.g., DeepMind's AlphaFold series). Commercially, these technological achievements have attracted unprecedented venture capital, with global AI funding surging past \$100 billion in 2024 \citep{glaser_state_2025}. Additionally, there have been high-profile strategic alliances between companies (e.g., Microsoft's deepened partnership with OpenAI and Google and Amazon's investments in Anthropic) to ensure market dominance in a hyper-competitive environment. Politically, competition with China has instilled a sense of urgency for Western innovation, which has been intensified by the release of high-capability open-source models (e.g., Meta's Llama series, xAI's Grok series) and the launch of the R1 model by the Chinese company DeepSeek in early 2025. In the United States, Donald Trump's election marked a pivotal moment. His administration swiftly replaced the previous safety-focused Executive Order with a directive aimed at ``removing barriers to American leadership in AI'' \citep{noauthor_removing_2025}. Moreover, it publicly backed an OpenAI-led investment of \$500 billion in AI \citep{noauthor_announcing_2025}. This proactive stance garnered international support at the AI Action Summit, hosted by France in February 2025. This summit solidified a shift in global consensus, redirecting AI governance discussions from a predominantly safety-based agenda to one that champions action, opportunity, and accelerated innovation.

\subsubsection{The Elite's Discursive Shift} 
The social context discussed in the previous section has influenced (and been influenced by) the discourse of the AI executive elite. This section examines potential social determinants contributing to the recent surge in pro-TSD discourse. This analysis is framed by the technological, commercial, and geopolitical forces outlined in the previous section.

First, the increasingly impressive and frequent technological breakthroughs of the last few years likely fueled a sense of awe and inspired visions of shared prosperity. Four of the top five (80\%) texts displaying Cult of Technology aspects were published within this period. Second, the AI `gold rush,' fueled by immense commercial interests, likely led to viewing AI as the next engine of economic growth. Technology companies have been rapidly adding AI capabilities to virtually all their products (cf. Tech Goggles, Solutionism), claiming significant gains in productivity. Four of the top five (80\%) Tech Goggles texts and the top five Solutionism texts were published within this period. Finally, the current hyper-competition has been fueled by the US administration's vocal support for American leadership in response to China's AI development. The ``bad actor'' threat is commonly used to justify many AI-related concerns. Three of the top five (60\%) texts were published within this period.

The determinants discussed thus far have likely contributed to a surge in the expression of core TSD tenets. However, the period following the launch of ChatGPT also saw a noticeable surge in TSD-reinforcing responses to AI-related concerns, particularly those devoted to marginalization. This surge can be primarily explained by \emph{discursive}, rather than technological, commercial, or geopolitical determinants. Ironically, the dominance of the safety-first discourse around the time ChatGPT was launched may have led to the marginalization of concerns. In early 2023, 18 out of 19 (95\%) elite members indicated their divergence from the dominant safety discourse by not signing the Future of Life Institute's open letter. Shortly after, this implicit stance became explicit opposition as many members started minimizing, dismissing, deflecting, or reframing AI-related concerns. Indeed, four of the top five (80\%) marginalizing texts were published within this period. The most extreme example of this overt conflict is \texttt{\small andreessen-2023-manifesto}, a manifesto with an explicit call to action against a perceived ``enemy,'' and the most voluminous source of dismissive discourse across the corpus. These observations suggest that discourse is not merely a result of its evolving social context, but a part of it, meaning that it can also be a powerful determinant for future discourse.

\section{Discussion}
\label{sec:discussion}

In response to objective I (how TSD is presented, challenged, or reproduced), this research demonstrates conclusively that the TSD discourse of the AI executive elite is not monolithic but reveals a broad spectrum of stances. While the average text tends to be moderately pro-TSD, this central tendency is marked by high variability, confirming that TSD is not a uniformly held belief system but a site of active discursive struggle. The analysis reveals that pro-TSD discourse is predominantly constructed through direct expressions of its core tenets, particularly visions of a utopian future and the framing of technology as the primary or inevitable driver of progress. Moreover, these two aspects of TSD are often combined, creating a powerful legitimizing narrative for the accelerated development and deployment of AI at scale. A key finding is the identification of Benign Techno-Optimism (BTO) and its distinct discursive profile. While displaying a genuine belief in AI's positive potential, BTO discourse is critically tempered by direct contradictions to core TSD tenets, particularly those related to Techno-Chauvinism, and explicitly values non-technological efforts. The distinction between full-fledged TSD and BTO ultimately revolves around their implicit theories of change. TSD posits that technology is the primary, autonomous driver of progress, and that society must adapt accordingly. In contrast, BTO frames technology as a powerful but contingent tool that must be actively and deliberately steered by humanistic values, governance, and social institutions to produce beneficial outcomes. The struggle between these two visions is not merely about optimism versus pessimism, but about where agency and ultimate control reside, in the technology itself or in the society that wields it. BTO's optimism is earned, not assumed; it coexists with a deep awareness of risks and an openness to diverse, non-technological mitigation strategies. While BTO may be less problematic than full-fledged TSD, it is crucial to critically question whether it could function as a more sophisticated form of legitimation that preserves the industry's central role under the more palatable guise of `responsible innovation.' BTO discourse could serve as a strategic shield, deflecting more radical critiques and ensuring the industry remains the primary architect of its own oversight.

Regarding objective II (response to AI-related concerns), the research finds that the most common form of critical engagement is not direct opposition to TSD core tenets but the acknowledgment of concerns. However, a crucial finding is that this acknowledgment often serves merely as a rhetorical preamble. It strategically validates concerns only to pivot to pro-TSD responses, most notably by proposing purely technological solutions or marginalizing the concern. This pattern reveals that while elite members frequently engage with risk, they rarely pose fundamental challenges to TSD itself.

In response to objective III (elite's discourse dynamics), the analysis reveals that the AI executive elite's discourse is not static but highly dynamic and adaptive. A key finding is the marked polarization of stances after the launch of ChatGPT, with most authors exhibiting a notable increase in pro-TSD discourse over time. This discursive shift seems to be a strategic adaptation to a rapidly evolving social context, moving from an initial focus on AI safety to action. This dynamic can be explained by a confluence of social determinants: impressive technological breakthroughs fueled the tenets of Techno-Chauvinism and the Cult of Technology; intense commercial competition promoted Solutionist and Tech Goggles framings; and geopolitical pressures provided justification for accelerated development. The increase in discourse marginalizing AI-related concerns seems to be a direct counter-narrative to the rising influence of the previously dominant safety-first discourse. This phenomenon positions the AI executive elite's discourse not merely as a reaction to the `AI gold rush,' but as a strategic intervention that actively works to construct, legitimize, and accelerate concrete visions for the future of AI.

\subsection{A Diagnostic Approach}

TSD (with its unfortunate anagram) can be thought of as an ideological contagion that spreads through discourse. Like a pathogen, its influence is not always immediately apparent. It can be challenging to detect because its tenets are often naturalized through repetition, eventually appearing as common sense. TSD can thus `infect' even the most well-intentioned efforts to tackle important issues, subtly shaping the framing of problems and limiting the range of considered solutions. Given its often subtle nature and the broad range of manifestations, developing a suitable diagnostic toolkit to identify TSD becomes crucial. The methodology developed for this paper functions like such a toolkit. It aims to move beyond intuition, offering a systematic methodology for detecting TSD, much as a trained medical professional uses specific tests to confirm a diagnosis rather than relying on general impressions. The quantitative metrics (TSD Adherence and Balance) provide quick tests that offer a rapid overview of a text, uncovering broad patterns and identifying outliers to articulate well-supported hypotheses. These quantitative insights inform the in-depth qualitative analyses that serve as more specialized diagnostic tests. These in-depth analyses examine specific linguistic `symptoms' to confirm the presence of TSD and understand its distinct manifestations (e.g., benign, insidious, overt, subtle). Just as an experienced doctor becomes progressively better at spotting diseases, TSD becomes easier to trace and critique once an analyst is sensitized to its discursive markers.

For CDA to contribute to social change, however, it must engage in dialogue with social actors who are in a position to act. The ultimate value of this paper lies in fostering awareness among key stakeholders. This awareness, however, must be critical and extend beyond simply being able to trace TSD. It is also crucial for these social actors to understand why TSD may be problematic and how its components reinforce one another. There are three social groups that must develop this critical awareness for TSD to be challenged effectively: AI practitioners, policymakers, and consumers. AI practitioners building and deploying applications (e.g., engineers, designers) must be able to diagnose TSD in their discourse and that of their leadership. Without this critical awareness, even well-intentioned efforts can fall into a techno-solutionist trap. By becoming sensitized to TSD, AI practitioners can question whether a technological fix is the best approach and implement genuine solutions that account for complex social contexts and integrate appropriate safeguards. Policymakers and regulators formulating governance frameworks (e.g., the EU AI Act) are subject to intense lobbying and sophisticated public discourse. Being able to detect the subtle workings of TSD (e.g., distinguishing a genuine acknowledgment of risk from a token gesture meant to preempt regulation) is essential for crafting effective, robust, and truly society-serving policies that are not captured by the very industry they seek to govern. AI consumers (i.e., virtually everyone with access to digital technology) play a crucial, often underestimated role. If consumers are sensitized to TSD, they can purposefully exercise their consumer and civic power to question assumptions and claims that often dominate marketing and public narratives. Deciding whether to use AI and which one should be conscious and intentional, informed by the consumer's own values.

Developing a critical awareness to trace TSD is necessary but insufficient for meaningful praxis. While the diagnostic toolkit developed in this paper empowers social actors to identify and support more responsible manifestations (e.g., Benign Techno-Optimism) over more insidious or aggressive ones, the final stage of critical engagement requires holding elite members accountable by verifying the congruence between their public discourse and their material actions. Accountability demands that we scrutinize whether a company's stated commitment to safety (e.g., Anthropic's Responsible Scaling Policy) is reflected in its product design choices; whether its advocacy for responsible AI aligns with its lobbying efforts on regulation; and whether its vision of shared prosperity matches its internal labor practices and investment priorities. Without this vital step of verifying alignment between word and deed, even the most seemingly responsible discourse risks functioning merely as a strategic shield for unaccountable power.

\section{Conclusion}
\label{sec:conclusion}

This paper's contributions should be contextualized by several limitations that offer potential avenues for future research. First, TSD may not be a consciously-held belief system. While this paper defines TSD as a cohesive ideology, its tenets might be more fragmented and/or opportunistic in practice. Future ethnographic or interview-based research could examine how well this framework aligns with the self-perceptions and motivations of elite members. Moreover, it could investigate the out-of-scope phenomena identified during the broad phase of this paper, including the role of saviorism given perceived moral motivations; internal disagreements and fragmentation within the elite (e.g., Andreessen versus Amodei on the merits of Universal Basic Income); and the quasi-religious nature of TSD discourse. A second limitation is the paper's scope. The findings may not generalize to the broader global AI ecosystem, given the focus on a systematically defined, US-centric executive elite. Future analyses could include a broader range of influential actors, including leaders of non-US companies (e.g., DeepSeek, Mistral) and key opinion formers (e.g., scientists, journalists). These analyses would create a comparative understanding of how TSD is reproduced or challenged globally and across different institutional roles. Additionally, the scope could be broadened to include shorter texts (e.g., tweets) and spoken discourse (e.g., podcasts), thereby allowing the inclusion of highly influential figures such as Demis Hassabis and Elon Musk. Another limitation is that this paper does not establish a causal relationship between elite discourse and societal outcomes. While premised on the principle that discourse has social effects, it cannot isolate or measure the actual impact of these discursive strategies in the real world. This limitation points to the need for empirical reception studies. Future projects could employ experimental methods or surveys to investigate how different manifestations of TSD, from benign to insidious, actually influence public opinion and/or policymakers' attitudes.

The most personally exciting and immediate future project is the development of an AI-powered TSD diagnostic tool. The underlying model would be fine-tuned on this research's annotations and findings and would be prompted to act as an expert in this paper's `T' methodology. The model could employ Retrieval-Augmented Generation (RAG) techniques to access TSD's and CDA's theoretical foundations and utilize Reinforcement Learning (RL) to incorporate user feedback on an ongoing basis. Such an application could analyze text, audio, and even video in near real-time, providing users with TSDA and TSDB scores, and identify concrete linguistic features that signal TSD's presence. To maximize its practical impact, the tool could position analyzed texts on the TSDA/TSDB spectrum relative to this paper's corpus and, crucially, link identified TSD aspects to established critiques and concrete examples of potential or actual harm. By presenting these insights in accessible formats (e.g., infographics, short videos, podcasts), this tool could translate critical academic research into a powerful instrument for public and professional awareness. This project would represent a reflexive application of AI, turning its analytical power back upon the ideological systems from which it emerges. This approach moves beyond a simple anti-technology stance to embody critical praxis. By leveraging AI's unique capabilities for scalable, real-time analysis, the goal is not to propose a technological `fix' for TSD, but to create an effective instrument for public sensitization.

This paper is an anti-TSD piece of discourse from a concerned techno-optimist. While technology can be a powerful aid in tackling complex societal issues, it must be employed consciously to prevent or mitigate harm equitably. The sheer velocity of AI development with its accompanying legitimizing discourse makes critical awareness an immediate necessity. While awareness alone does not constitute transformative action, it is an essential prerequisite. By uncovering the Techno-Supremacy Doctrine's core components, discussing its risks, and providing a diagnostic toolkit for its manifestations, this paper aims to foster a collective sensitization among AI practitioners, policymakers, and consumers. We all have the responsibility to act. Let us support or challenge dominant discourses both consciously and materially, demanding that the promises made are met with accountable action. Let us navigate the future of AI as active participants, not passive recipients of an ``inevitable'' destiny manufactured unilaterally by the AI executive elite.

\bibliographystyle{unsrtnat}
\bibliography{references}

\end{document}